# Kolmogorov´s Legacy: Algorithmic Theory of Informatics and Kolmogorov Programmable Technology

**Sergei Levashkin**

Artificial Intelligence Lab

sergei@levashkin.com

**Victor Alexandrov**

Russian Academy of Sciences, Saint Petersburg, Russia

alexandr@iias.spb.su

**Adolfo Guzmán-Arenas**

Instituto Politécnico Nacional, Mexico City, Mexico

aguzman@ieee.org

**Abstract.** In this survey, we explore Andrei Nikolayevich Kolmogorov's seminal work in just one of his many facets: its influence Computer Science especially his viewpoint of what herein we call "Algorithmic Theory of Informatics."

Can a computer file "reduce" its "size" if we add to it new symbol(s)? Do equations of state like $2^{nd}$ Newton law in Physics exist in Computer Science? Can Leibniz' principle of identification by indistinguishability be formalized?

In the computer, there are no coordinates, no distances, and no dimensions; most of traditional mathematical approaches do not work. The computer processes *finite* binary sequences i.e. the sequences of 0 and 1. A natural question arises: Should we continue today, as we have done for many years, to approach Computer Science problems by using classical mathematical apparatus such as "mathematical modeling"? The first who drew attention to this question and gave insightful answers to it was Kolmogorov in 1960s. Kolmogorov's empirical postulate about existence of a program that translates "*a natural number into its (binary) record and the record into the number*" formulated in 1958 represents a hint of Kolmogorov's approach to Computer Science.

Following his ideas, we interpret Kolmogorov algorithm, Kolmogorov machine, and Kolmogorov complexity in the context of modern information technologies showing that they essentially represent fundamental elements of Algorithmic Theory of Informatics, Kolmogorov Programmable Technology, and new Komputer Mathematics i.e. Mathematics of computers.

# 1    Introduction

In the present work, we embed into the context of modern information technologies and their applications the contributions of the greatest Soviet mathematician Andrei Niko-layevich Kolmogorov (Андрей Николаевич Колмогóров, April 25, 1903, October 20, 1987) to Computer Science. Kolmogorov is well known in many sciences: Mathemat-ics, Celestial and Classic Mechanics, Physics, Genetics, Linguistics and many others where he pioneered new theories and made fundamental discoveries. In our opinion, his work on Programmable Technology and Algorithmic Theory of Informatics is much less known, especially among IT-practitioners. That is why this survey will thoroughly follow Kolmogorov's legacy in algorithmic theory of informatics[1], theory of algorithms and machines, programmable technology, and Komputer mathematics. In Informatics, Kolmogorov is best known by his work on algorithmic complexity. From our point of view, K-complexity can be seen as an auxiliary concept of software effectiveness. Nev-ertheless, it is a central concept in Computer Science, since it links programs, algo-rithms and their representation or encoding as binary strings. However, the Kolmogo-rov's legacy in Computer Science is much more than K-complexity. In this article, we try to justify this thesis and draw the attention of the Computer Science community (practitioners first) to the richest Kolmogorov's heritage in the field. Note also that the famous uncomputability[2] of K-complexity significantly reduces the interests of practi-tioners in Kolmogorov's theories, leaving them in the hands of mathematicians. Huge gap! We try to fill this gap too, that is another goal of present publication. In our text we follow (more or less chronologically) the order of appearance of Kolmogorov's most important works [8], [9], and [10] aiming to show the logics of their evolution which finally led Kolmogorov to visionary concepts of new K-mathematics[3]. This is in fact very typical for Kolmogorov's creativity: Starting with a difficult but particular problem and then obtaining far general results. The same occurred, for example, with the 13th Hilbert problem: The methods used by Kolmogorov's student Vladimir Arnold to solve the original problem in Hilbert's statement were generalized to solve much more general problems by Kolmogorov himself [7]. It is interesting to note that in es-sence [8]-[10] were the last "serious" Kolmogorov scientific publications. Then he en-tered a period of clarification of some his earlier works, and writing textbooks in math-ematics for mid and high schools. By the way, the writing of textbooks is a titanic task, just at the scale of Kolmogorov's personality. However, Kolmogorov in his career did not take simple tasks. The enormity of solved problems can most clearly be illustrated by Kolmogorov algorithmic theory of informatics and, based on it, Kolmogorov programmable technology (see Section 3 in the following).

---

[1] In the present paper, we use the notions "Informatics" and "Computer Science" depending on current con-text. However, we understand both in their common sense as they are defined in widely accepted ency-clopedias of Computer Science, for example. Alternatively, one can think about somewhat like Infomat-iCS that is Informatics and/or Computer Science. We also use the notion Komputer Mathematics or K-mathematics instead of Computer Mathematics to distinguish Kolmogorov's approach to the mathemati-cal problems of Computer Science (see Section 3 for accurate definitions).

[2] In fact, the computability of K-complexity is not so relevant in applications of K-technology. See in the following Section 3.

[3] Along the paper, we also use the abbreviations ATI or @I for Algorithmic Theory of InformatiCS, KPT or K-technology for Kolmogorov Programmable Technology; K-postulate for Kolmogorov postulate; K-machine for Kolmogorov machine; K-complexity for Kolmogorov complexity; K-algorithm for Kolmo-gorov algorithm (see Section 3).

The rest of the article is organized as follows. In Section 2, we describe several previous and current works, to embed the objectives of the article into the context of the state- of- the- art. Section 3 contains the main contributions of the article. We introduce and explain the original concepts of Computer Science that belong to Kolmogorov. We follow the works of Kolmogorov in chronological order and the concepts described in these works as he introduced and developed them. These concepts are:

- Kolmogorov postulate (K-postulate, §3.1), which refers to deep formalization of Leibniz' principle identification by indistinguishability;
- Kolmogorov algorithm (K-algorithm, §3.1);
- Kolmogorov complexity (K-complexity, §3.1 and §3.3);
- Kolmogorov machine (K-machine, §3.2).

They represent the theoretical basis of the

- Kolmogorov Programmable Technology (KPT or K-technology, §3.1 and §3.5) and the foundations of
- Komputer mathematics or K-mathematics (§3.4).

## 2  Previous and current work

What approaches preceded the emergence of Kolmogorov's theories? What are current trends related to the Kolmogorov's approach? In this section, we outline some of them.

a) There are many definitions of the notion "information". From the most common sublimely philosophical such as "information is a reflection of the real world" or "information is a general property of matter and the measure of organization of systems" to a purely practical: "information is data that is subject to the collection, conversion, storage, and transmission". Anyway, information theory deals with mapping objects or phenomena into symbols, or images. Symbols can be very diverse as, for example, a sequence of electro-magnetic impulses coming from the telecommunication satellites, spoken and written language, TV pictures, genetic code that contains the inherited properties in biological cells. *Finding an optimal system of symbols that describe the properties of objects (encoding) and obtaining data about objects based on the properties of symbols (decoding) are the typical problems of the theory of information.*

b) In the introduction to the first edition of the book "Cybernetics" [17], which was published in 1948, Norbert Wiener wrote: ... *we had to develop a statistical theory of the amount of information. In this theory, the amount of information per unit is the quantity transferred in a single choice between equally probable alternatives. This idea emerged almost simultaneously from several authors, including a statistician R.A. Fisher, Dr. Shannon from Bell Telephone Laboratories, and the author of this book. Wherein R.A. Fisher proceeded* *from the classical statistical theory, Claude Shannon from the problem of information encoding, the author of this book from the problem of communication and noises in electrical filters. It should be noted that some of my research in this area is related to the earlier work of Kolmogorov in Russia; although a large part of my work was done before I turned to the works of the Russian school.*

c) Shannon's work [16] was published in 1948. Using the methods of probability theory**,** Shannon sought for the best ways of encoding and decoding information for its transmission and storage. The first publication of Shannon, written at the level of "physical rigor", attracted the attention of Kolmogorov. In the preface to the Russian translation of these works, Kolmogorov wrote: *The importance of Shannon's works for pure mathematics was not immediately appreciated enough. I recall that even at the International Congress of Mathematicians in Amsterdam (1954) my American colleagues, specialists in probability theory, thought my interest in the work of Shannon somewhat exaggerated, as it is more technique than mathematics. Now such views are unlikely to need refutation. However, a rigorous mathematical "justification" of his ideas Shannon in any difficult cases provided to its successors. Anyway, his mathematical intuition is amazingly accurate...*[4].

d) *Morse code.* An illustration of the center point idea of Kolmogorov's algorithmic theory of informatics *a natural number into its binary record and the record into the number* (see Section 3.1 in the following) is Morse code (1836) [5]. This optimal non-equilibrium code takes into account the rate of the most frequently used letters in the text. It is a method of sending text messages by keying in a series of electronic pulses, usually represented as a short pulse (called a "dot") and a long pulse (a "dash"). A kind of Morse code is Barcode developed by Norman Woodland *et al.* (1949) [4]. In barcode, a dot is a "thin bar" and a dash – "fat bar". Both codes follows the same idea of information content quantification; both codes deal with information objects and their identification. Kolmogorov formally developed precisely this idea: first at empirical level (*postulate*) [8] and then formally (*algorithmic theory of informatics* and *programmable technology*) [8] which finally led him to the foundations of *Komputer mathematics* [10].

e) Arnold, using Newtonian differences and Leibnizian monads, proposed an interesting interpretation of the structural complexity of finite binary sequences [3]. His approach involves the construction of the graph corresponding to a given binary sequence and study its properties serving as a measure of (structural) complexity of the bit string by using the geometric properties of the graph. The sequence is considered more complex if it belongs to a cycle of the graph that has greater length. If two sequences have the same cycle length, then the sequence most distanced from the cycle is more complex.

---

[4] That is the whole Kolmogorov as it is! On the one hand, "physical level of rigor" and on the other hand, "his mathematical intuition is amazingly accurate". Namely**,** Kolmogorov and his disciples were who brought the results of Shannon to rigorous mathematical justification in any difficult cases. Shannon got in their face worthy successors.

f) Manin and Marcolli [14] interpret K-complexity as a kind of energy similar to the energy of thermodynamic (chaotic) systems. The origin of their approach is Landauer's principle.

g) Ont of the most complete interpretations of Kolmogorov approaches to algorithmic theory of informatics can be found in Alexandrov *et al.* [2] and Levashkin-Alexandrov [11]. In [2], for example, its authors proposed an equation of state that links information and energy in mobile phones in terms of K-complexity. Due to the fact that GSM phones transmitted data in a digital format, the battery life depends on the number of transmitted bits: $P = klS(I)/t$, where $P$ – power consumption ($W$); $l$ – length of the information message ($bit$); $t$ – time of transmission ($s$); $S (I)$ – a function of the complexity of the information message indicating how many times the message is compressed; $k$ – a dimensional constant ($W \cdot s / bit$). *Probably, to date this equation is the first and only equation of state in Computer Science*! Then applying Einstein's formula $E=mc^2$, they obtained from the previous formula that $I \leq KE = Kmc^2$, where $E$ – energy necessary for transmitting a message in $I$ bits; $K$ – is Kolmogorov coefficient showing the effectiveness of the best selected program: formats and protocols, data storage and transmission as well as the level of technology and efficiency of equipment. The smaller (closer to 1) is $K$, the closer to optimal is the transmission technology. The authors interpret this formula as the minimum weight of a carrier that holds a desired amount of data, or equivalently – minimum energy required for transmission of the same data volume. Then the required transmission power of the message is $PS$ Watts at a speed equal to $S$ bit / s, if for the transmission of one bit of data is required to expend energy in $P$ Joules (given transmitter power and its auxiliary power units). Consequently, compression algorithms reduce the length of a message at a lower bit rate, which decreases the required power.

*Remark*. In [11], a preliminary version of the present article, its authors interpret Kolmogorov's theories in a similar way.

The first major task of the new theory was to find a quantitative measure of information, i.e. numerical estimation of messages "informativeness." In developing this idea, i.e. characterizing uncertainty message sources, Shannon used the notion "entropy" [16]. The entropy allows solving many important problems related to the transmission and storage of digital messages. However, all attempts to generalize it to the case of continuous signals were unsuccessful. *Then I insist on the idea that the basic concept, which can be generalized to completely arbitrary continuous messages and signals is not directly the concept of entropy, but the notion of the amount of information I (x, h) in a random object x with respect to the object h.* This phrase from the Kolmogorov's report *Theory of Information Transmission* at a meeting of the Academy of Sciences of the USSR in 1956 and the following formal mathematical constructions identified a new algorithmic theory of informatics [11].

## 3    Kolmogorov Algorithmic Theory and Technology

From the late of the 1950s until the beginning of the 1970s, Kolmogorov reconstructed the foundations of information theory. About this period of his career, he wrote: ... *my quite common, semi-philosophical thoughts have taken more time and energy than it can be seen from afar. In such attempts to formulate very general views up the outcome of the efforts is not exactly in wording fixed "results", but in a total restructuring of own consciousness and placing all ideas in the proper perspective. Therefore, then it appears that, as it were, you discovered nothing "new", but spent a lot of time and effort.* Kolmogorov is disingenuous here. He discovered "only", a new trend in information theory. Kolmogorov's articles [8]-[10] gave birth to the algorithmic theory of informatics, programmable technology, and new Komputer mathematics.

In the rest of this section we outline these Kolmogorov achievements, starting from Kolmogorov's work [8] done with his student Vladimir Uspenski (Sections 3.1 and 3.2), where the concepts of K-algorithm and K-machine were introduced. Section 3.3 is about K-complexity [9], where the reader finds its unusual interpretation. We consider in Section 3.4 Kolmogorov's work [10] where he drew a sketch of Komputer mathematics. We close this part with a brief overview of some modern applications of the Kolmogorov theories (Section 3.5).

### 3.1    Algorithmic Theory of Informatics and Kolmogorov
###           Programmable Technology

In [8] Kolmogorov and Uspenski introduced the most general to date definition of algorithm (let us call such algorithm – K-algorithm). Our point is that the concepts of K-algorithm, K-machine, and K-complexity just interpreted in terms of modern information technology (IT) represent a deepest and most advanced IT theory – algorithmic theory of informatics.

In 1950-1960 Kolmogorov and other mathematicians looked for a more rigorous definition of the algorithm most adequate to the functioning of computers. Very soon, it became apparent that the concept of the algorithm rooted as Al-Khwarizmi "arithmetic" has nothing to do with the concept of the algorithm as "computable function", implemented by Turing machine.

**K-postulate.** A link between the concepts of "process" and "computable function" stemming from the structural, constructive organization of the computational process formulated by Turing in 1936 and named "Turing machine" was deeply rethought by Kolmogorov and formulated in a footnote of the article [8] as the following empirical postulate (*K-postulate*) that also contains the idea of programmable technology:

*… A method allowing to find the number[5] of a record[6] and to restore the record itself by its number is typically quite simple (so that the existence of an algorithm[7], "processing" the record into its number, and the algorithm, "processing" the number into its record is beyond doubt).*

For programmable technology, another important and very non-trivial to the 1960s, was a Kolmogorov reasoning about the representation of data: ... *the standard way of specifying information is by binary sequences, which start with units 1, 10, 11, 100, 101, 110, 111, 1000, 1001, ...;* **they are binary records of natural numbers**. *We denote by l (n) a sequence of length n.*

*Let we deal with a domain D of objects, which already has some standard numbering of objects by numbers n (x). However, explicit record of the number n (x) is not always the most "economic" way to find an object x. For example, the binary record of the number 9^9^9^9 is immensely long, but we can algorithmically define it very simply as 9 in 9 in 9 in 9.*

**From K-algorithm to K-complexity.** We first need the comparative study of different methods of specifying the objects of D. It suffices to consider only those methods that, to each binary record of the number p, a number – a unique identifier of information content $n = S(p)$ is assigned.

Thus, the method of specifying an object of D becomes nothing more than a function S of natural argument with natural values. A little further, we turn to the case where this function is computable. Such methods of setting S can be called "effective". However, we will maintain a full generality. For each object x of D it is natural to consider leading thereto p of smallest length $l(p)$. This shortest length will be the "complexity" of the object x for "method of setting S": $K_s(x) = \min l(p)$, $S(p) = n(x)$.

In terms of Informatics, p can be called "program" and S – "programming method"[8]. Then we can say that $l(p)$ is the minimum length of the program in which the object can be obtained using the method of programming S.

**From K-complexity to K-technology.** It should be emphasized that Kolmogorov replaced the concept of "algorithm" with the notion of "program" in his definition of K-complexity[9] and regretted that this fundamental result was not clearly seen and sufficiently thought by neither theoreticians nor practitioners. In fact, these are the main concepts of modern **information technology** – invariant representation of the binary sequence of data, carrying any kind of information content and **programmable technology** – instruction programming for reproduction of this data.

More formally. Among computable functions, $S(p)$ there is an optimum. For any computable function $K_s(x) \leq K_{s'}(x) + l(S, S)$. Essentially, $K_s(x)$ is a criterion for evaluation of the software conciseness that determines its effectiveness in terms of either required volume for data transmission, or velocity of the deployment procedure of terminal program, format, and narrative code into the data [2], or required size of the storage holding it. *Kolmogorov was first to draw attention to the specifics of Komputer mathematics, formally stating the foundations of **programmable technology, algorithmic theory of informatics, and not just algorithmic theory of information**.*

---

[5] natural number

[6] binary record

[7] K-algorihtm

[8] A programming method could be, for instance, the use of a given programming language.

[9] Obviously, there is an algorithm beyond any program.

Kolmogorov emphasized that his approach to information is based on object identification (the amount of information in an object is inherent in the object itself), rather than properties of the ensemble of objects which usually are transmitted in addition to the given object (Shannon). This meaning is much closer to the modern understanding of Computer Science/Informatics.

We underline that the concepts introduced by Kolmogorov (natural methods of programming, complexity, and software quality in terms of complexity) do not depend on a particular programming language, they are universal.

### 3.2    K-machine

Working on problems of algorithmic theory of information and general theory of algorithms, Kolmogorov introduced the notion of machine now known as Kolmogorov or K-machine. In some sense, it was an auxiliary theoretical concept for new Komputer mathematics and an alternative for Turing or T-machine. The main difference between K and T machines is that the tape of K-machine can change its topology while the tape of T-machine cannot. Essentially, the tape of K-machine is a finite connected graph with a distinguished (active) node. The graph is directed but symmetric: If there is an edge from $u$ to $v$ then there is an edge from $v$ to $u$. The edges are colored in such a way that all edges coming out from a node have the same color, but each node colors its outgoing edges with a different color. The number of colors is bounded (for each machine). It is clear from the definition (*the tape is a graph*) that Kolmogorov thought about structural (*semantic*) representation of machine states! See details and formal definitions in [8]. Nevertheless, the most surprising result is that K-machines in certain sense are more powerful than T-machines! Namely, the following theorem is true: *There is a function real-time computable by some K-machine but not real-time computable by any T-machine* [6].

### 3.3    K-complexity: Three approaches

The last in the series of Kolmogorov's concepts related to information theory and Computer Science is K-complexity. The problem of defining the complexity of an object is a fundamental and one of the oldest scientific problems. In addition, it should be noted that this problem is one of the most difficult to formalize. Indeed, solving it, one must answer the question: Given an object, what is a measure of its complexity? In other words: Which objects are "complex", which are "simple", and which are "average"? So, which objects are considered complex?

Kolmogorov gave a surprising answer to this question by introducing K-complexity. In [9], literally on ten pages of text in a clear manner, he summed up the decades of numerous studies of the problem to define "the amount of information" $I(x)$. He identified and described three main approaches to the problem: *combinatorial*, *probabilistic*, and *algorithmic*. In the latter case, he gave the above definition of K-complexity and deeply studied its basic properties. Furthermore, he also gave a definition of relative complexity $K(x,y)$ – the complexity of the object $x$ somehow related to the other object $y$ and the relative amount of information $I(x,y)$. The main conclusion of the article is paradoxical: *Information theory must precede the theory of probability, not rely on it*. And this said the man who transformed the theory of probability, which was for centuries the games of dice and cards, into formal, mathematical science! Note also that this thesis marks the fundamental difference between the approaches of Kolmogorov and Shannon because Shannon based his theory on probability i.e. the probability for him precedes the information and not vice versa. Therefore, following Shannon we have to focus on statistical processing while following Kolmogorov on semantic processing.

Science has its own, internal logic and it is unlikely that this definition would be possible without the appearance of the first computers. Let us illustrate this thesis.

1. By the time of appearance of Kolmogorov's definition it became clear (largely due to the works of Shannon) that the most convenient and implementable way of describing any object is binary sequences (i.e. finite sequences of 0 and 1).
2. Binary objects comfortably journey through the communication channels, while obeying certain laws discovered by Shannon.
3. The advent of computers has led to the rapid development of programming tools and programming languages in the early to mid-1960s. Moreover, Turing, Universal Turing, and Kolmogorov Machine provided broad theoretical basis for development of programming tools and Computer Science in general.

If we look at the definition of Kolmogorov, it becomes clear that it is a surprisingly compact, elegant, and natural combination of the above-described circumstances. Again, science has its own, internal logic of development and Kolmogorov in his long-life work, as no one else, was able to reflect this logic. In our understanding, algorithmic theory of informatics and programmable technology, whose ideas most clearly appeared in the Kolmogorov's presentation at the International Congress of Mathematicians in Nice (1970) [10] became a worthy conclusion of Kolmogorov's brilliant scientific career. In this article, we focus on the latter, and not so much on K-complexity [11].

### 3.4    K-mathematics

We recall that the concept of the algorithm rooted as Al-Khwarizmi "arithmetic" has nothing to do with the concept of the algorithm as "computable function", implemented by Turing machine. For instance, 7+5=12 is a trivial problem for the Al-Khwarizmi arithmetic and very non-trivial for the computing. In the computer, "7" "plus" "5" is "equal to" "12" is true iff the numbers 5, 7, 12, the operation "+", and the equality "=" are defined, and can be "proved" only executing a program based on an algorithm, i.e. using programmable or K-technology.

As we saw above, Kolmogorov's approaches do not depend on any particular programming language. However, "the art of programming" is still focused only on translation from the formalism of the problem statement language to a limited basis of machine instructions that implement a given axiomatic (arithmetic) manner of data processing based on the principles of mathematical modeling and functional analysis.

**Object identification.** Historically, the first instruments of people's informative communication were natural language, art, and music. Much later, the mathematicians introduced a number and algebraic structures. Aristotle introduced logic. However, logic brings the problem of the consistency of the statements, and with it, the ambiguity of the object identification. After that, an idea of information exchange appeared among mathematicians. Thus, a "second mathematics" – the exchange of objects, which dates back to the Cantor's set theory and the Leibniz' principle of identification by indistinguishability, was born [12]. *Kolmogorov linked these processes (exchange of objects and identification) into the concept of algorithmic theory.* Namely, an interpretation of the algorithmic theory leads to the concept of the algorithm, which includes input data and process (instructions) to obtain a result. It comes down to the issue of the identification problem, i.e. the operation of translation of the result identifier by the input data identifier. This is explicitly implemented in the programmable logic blocks FPGAs and IP-addressing [12]. *In the near future, with the development of information technologies, we only have to deal with information objects, i.e. with KPT.*

**K-technology and Gadgets.** Components of digital communication systems, using the logic of Aristotle as a formalism, inherited the problem of identification, which for the human brain is not so relevant. In addition, the problem of digital processing closure, which came with algebraic structures, contributes nothing but contradictions. For example, the division of integer on an integer gives a floating point number (float), i.e. processing operations are not closed. Some mathematical axioms are unenforceable in digital processors and, consequently, in computers and controllers based on them. The digital implementation of most of these techniques goes through a series of useless operations to transform the problem in terms of mathematical description and software development of this description. In turn, the computer only emulates "mathematical" numbers processing at the software, but not at the hardware level. This means that *the physical processes occurring in the hardware of the computer and even their logical description do not correspond to mathematical numbers processing.* The processor operates with bits and bit sets. These sets of bits are identifiers (symbols, pointers) or numbers themselves. This means that the pointer should not necessarily be interpreted as a number. For pointers, it is sufficient to have the property of uniqueness to provide identification.

These and many other similar and well-known problems of classical mathematical modeling when it is applied in Computer Science emerge with the development of new – Komputer mathematics. Kolmogorov clearly understood this: *the heat equation in partial derivatives is equally far from the real physical process of heat propagation as its discrete model,* he wrote in [10]. In his works from the late 1950s to the late 1960s, which culminated with Kolmogorov's presentation at the International Congress of Mathematicians in Nice (1970) [10], he rethought and reconstructed the building of classical mathematics, stating in particular that: *The future is for discrete and Komputer mathematics!*

**Machine learning applications for differential equations.** KPT can be applied to solve initial and boundary value problems for high dimensional nonlinear differential equations. These problems represent hard analytical and computing difficulties by using classical approaches leaving many of them unsolved. While using algorithmic approach the problems can be solved as follows. A test solution of the differential equation is written as a sum of two parts. The first part satisfies the initial/boundary conditions and contains no adjustable parameters. The second part is constructed to not affect the initial/boundary conditions. This part involves a feedforward artificial neural network containing adjustable parameters (the weights). Thus, by construction, the initial/boundary conditions are satisfied and the network is trained (using algorithms of machine learning) to satisfy only the differential equation. With the advance of neuro-processors and digital signal processors, the method becomes very powerful due to the expected essential gains in the execution speed and exactness of solution [18].

### 3.5 Some modern applications of Kolmogorov Programmable Technology

KPT opened ways for development different kinds of software: compressors, codecs for fast convolution and involution (sweep) of bit sequences of the information content. At the same time, strong limitations on bandwidth, spectral, and energy characteristics are replaced by the ever-increasing, volume-speed characteristics ((kilo, mega, giga, tera, peta...) bits / s) of data processing and transmission. KPT still only at the *empirical level* plays a fundamental role in the development of various software products: Operating systems – Windows, OS, Linux; compressors – RAR and ZIP; tools – Photoshop, Flash, JPEG-codecs, MP3-players, etc.

**Mobile phone.** Let us suppose that two men speak by mobile phones $A$ and $B$. $A$ transmits to $B$ a program of compressed audio file instead of audio stream itself, while $B$ has the decoder. The program is very small. The decoder is very fast. Therefore, this kind of telecommunication is much more optimal than in common state-of-the art mobile phones (GSM alternative). Moreover, the program can be adapted to the individual features of the men's speech.

**Algorithmic compressor.** From non-monotonicity of K-complexity there follows a quite interesting and non-obvious observation: It would seem that by attributing to a file new symbols, the size of its archive should not be reduced; however, understanding "reduced" asymptotically it proves to be true! In fact, if large file was encoded shorter than small, the small one can be encoded, showing a large file code and adding to it a piece of code (of constant size), "chopping off" the extra characters.

**Information security.** Most of "random" number generators like "white noise" have very low K-complexity (the length of the program that produces them is quite small). Indeed, a "truly" *random sequence by Kolmogorov has K-complexity greater than the sequence length*. Moreover, a very short program can simulate it and therefore "random" white noise is in fact deterministic! In problems such as cryptography or, say, online gambling, where we want unpredictable results of the generator and the inability to predict the next result, high Kolmogorov complexity of a random sequence is critical, and in this case it may be a rough estimation of the effort required by a hacker to learn predicting the future behavior of the system. Thus, the conventional linear congruential generator may not be appropriate, since the hacker can enumerate several standard algorithms and a set of initial conditions for them having a non-zero chance (this chance can never come, of course) to find a combination that reproduces the system behavior. A better idea is to replace pseudo-random generators like white noise by truly random generators like "move arbitrarily your computer mouse a few seconds over table surface".

**Softwaring.** The main humankind activity of the past half century is "softwaring": *Absolutely anything that can be deprived of its physical body or physical incarnation will be softwared i.e. converted into a computer program*. Let us remember some examples. The shortest life cycle from an individual physical object to the icon has done by the pager – a single device has become a SMS on your phone in just two years. iPod took slightly longer time to be softwared. Initially it was a device with a spinning wheel (A). Then the wheel was painted – the moving parts are gone, the rotation sound came from the speaker (B). Then it turned into a metaphor of the wheel, and then – an icon (C, D). It is finally left and now it is just a program "Music" on your smart phone.

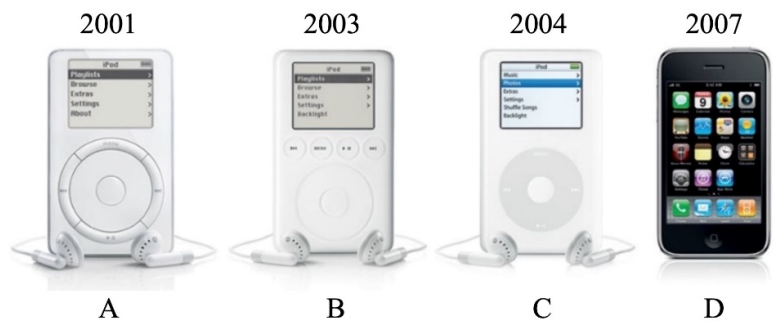

2001      2003      2004      2007

A      B      C      D

**Fig. 1.** iPod softwaring. A/ the wheel and bottoms are physical (2001); B/ the wheel has been softwared, the bottoms are physical (2003); C/ the wheel and bottoms have been softwared (2007); D/ iPod now is just an icon of the iPhone (2007).

Kolmogorov standing at the very beginning of the époque of the softwaring intuitively understood and foreseen this process by introducing K-complexity and thus putting the programming/softwaring as the keystone of the coming up technological age. That is the essence of K-technology described herein: *All that can be softwared will be softwared!*

K-technology can also be applied in the hottest problems of today such as bioinformatics and text mining [11], [2]. A vast majority of emerging seemingly alternative Computer Science theories and approaches are attempts to correct the fundamental limitations of functional analysis in their computer simulation. These approaches are still in the stage of development of different formats, protocols and programs when used to measure some property of an individual object, data from the assembly of objects to which the individual belongs, instead of using only the features of that individual. While, according to Kolmogorov, the amount of information is the length of the "program", when machine reproduces the binary data sequence of information content. Thus, adapting Kolmogorov's approach we are getting closer to the emergent needs of Computer Science applications as described in this Section.

## 4    Conclusion

K-algorithm, K-complexity, and K-machine are three elephants on whose backs the algorithmic theory of informatics is standing. Kolmogorov thoroughly revised the theory of information, extending it to very general concepts, which, from our point of view, represent mathematical foundations of modern digital technologies.

Moreover, in the work [3] Kolmogorov formulated his understanding of mathematical principles that we should follow in Computer Science. Essentially, they were the foundations of new mathematics that we call herein "Komputer mathematics". Let us recall some key points. It is well known that some mathematical axioms are not fulfilled in the computer. For example, the results of computing $z \cdot (x/y)$ and $(z/y) \cdot x$ for the computer, in general, are not equivalent. For the computer there is not much difference between countable and uncountable sets, because representation of real numbers is discrete. In the computer, theorems on limits, integral and differential calculus and so on are incorrect. All of this is a consequence of the artificial transfer of arithmetic as "mathematical basis" to computer processors running on the "logical basis" [15]. That is why in [1], for example, a possibility of replacing the computing "arithmetic" of mathematical model by its direct translation into processor commands is considered. In other words, not a mathematical model is embedded into the "Procrustean bed" of the computer, but on the contrary, its computer representation (memory, CPU/GPU/TPU) and its axiomatic basis (commands) should be formalized.

Kolmogorov also formulated principles of programmable technology (KPT) by introducing the concept of complexity as a criterion for evaluating the effectiveness of the program in terms of its length. Furthermore, in his widespread bibliography this **auxiliary concept** is defined as "theory of complexity of constructive objects" instead of the global priority in the development of *algorithmic theory of informatics* and the foundations of *programmable technology*. This is a clear and unfortunate example of what happens when scientific intellectual property is not recognized and rarely assigned at the stage of technological embodiment[10].

Summing up. The very rich Kolmogorov heritage urgently requires careful study and practical development in the applications of the ultimate information technology instead of "sport competitions" between mathematicians [19]. We have no doubt that this legacy will certainly contribute to IT further improvement and development.

### Acknowledgments

S. Levashkin wishes to thank A. Kolmogorov who was his teacher in the Moscow High School #18.

V. Alexandrov thanks the co-authors of this paper for very fruitful collaboration.

A. Guzman appreciates the partial support of CONACYT, SNI and SIP-IPN.

Note: A reduced version of this work appeared in Spanish [13], for the benefit of the Spanish-speaking computer science community, thanks to IEEE Section 9.

---

[10] This is one of the essential differences of our interpretation from others.

[11] Originally published in the journal "Russian Mathematical Surveys", 1957.

[12] Originally published in the journal "Russian Mathematical Surveys", 1958.

---

[13] Originally published in the journal "Problems of Information Transmission", 1965.

[14] Originally published in the journal "Russian Mathematical Surveys", 1983.